\documentstyle[12pt,aaspp4]{article}

\def\ltsima{$\; \buildrel < \over \sim \;$}
\def\simlt{\lower.5ex\hbox{\ltsima}}
\def\gtsima{$\; \buildrel > \over \sim \;$}
\def\simgt{\lower.5ex\hbox{\gtsima}}

\journalid{}{}
\articleid{}{}
\newcommand{\Ref}{\hangindent=20pt \hangafter=1 \noindent}
\newcommand{\StartRef}{\hyphenpenalty=10000 \raggedright \parskip=0pt \parindent=0pt }
\righthead{Dusty Ly$\alpha$ Emitters at High Redshift}
 
\begin{document}
 
\title{Models for Dusty Ly$\alpha$ Emitters at High Redshift}

\author{Zolt\'an Haiman,\altaffilmark{1,2} \& Marco Spaans,\altaffilmark{2,3}}

\altaffiltext{1}{NASA/Fermilab Astrophysics Center, Fermi 
National Accelerator Laboratory, Batavia, IL 60510, USA}
\altaffiltext{2}{Harvard Smithsonian Center for Astrophysics,
60 Garden Street, Cambridge, MA 02138, USA}
\altaffiltext{3}{Hubble Fellow}

\begin{abstract}

Models are presented for the Ly$\alpha$ emission of dusty
high--redshift galaxies by combining the Press--Schechter formalism
with a treatment of the inhomogeneous dust distribution inside
galaxies.  It is found that the amount of Ly$\alpha$ radiation
escaping from the galaxies strongly depends on the time over which the
dust is produced through stellar activity, and on the ambient
inhomogeneity of the HII regions that surround the ionizing OB stars.
Good agreement is found with recent observations, as well as previous
non--detections.  Our models indicate that the dust content builds up
in no more than $\sim 5\times 10^8$ yr, the galactic HII regions are
inhomogeneous with a cloud covering factor of order unity, and the
overall star formation efficiency is at least $\sim5\%$.  It is
predicted that future observations can detect these Ly$\alpha$
galaxies upto redshifts of $\sim8$.

\end{abstract}
 

\vfil\eject

\section{Introduction}

One expects that the ionizing radiation from young stars leads to 
prominent Ly$\alpha$ emission through the recombination of hydrogen in 
the interstellar medium (Meier 1976).  Because the Ly$\alpha$ line is 
narrow and strong, it stands out against the continuum background and 
should provide a signature of primeval high--redshift galaxies 
(Partridge \& Peebles 1967).  Until a decade ago, the search for these 
high--redshift Ly$\alpha$ galaxies had enjoyed no compelling successes 
(e.g.\ Djorgovski 1992; Thompson et al.\ 1995).  A large class of high 
redshift radio galaxies was discovered in recent years (c.f.\ Chambers 
\& Miley 1990), but their numbers are still significantly below the 
theoretical predictions of about $10^{3-5}$ of such primeval galaxies 
per square degree (cf.\ Pritchet 1994).  As was pointed out in the 
above paper, the lack of detections was becoming a source of confusion 
and concern.  However, in the last couple of years, with improved 
sensitivity on large area telescopes, these young galaxies are finally 
being detected (e.g.\ Hu, Cowie, \& McMahon 1998; and references 
therein).  Clearly, this population is of great interest to the field 
of galaxy formation and the early evolution of the universe.  It is 
mandatory to understand why earlier surveys have been 
unsuccessful, how many objects are still expected to be found, and 
what physical conditions pertain in these early systems so that the 
Ly$\alpha$ radiation may escape.

On galactic scales, the spatial distribution and the amount of stellar 
dust are crucial ingredients for the radiative transfer effects in the 
Ly$\alpha$ line, such as resonant scattering and dust attenuation, and 
will determine the emerging Ly$\alpha$ luminosity (e.g.\ Neufeld 1991).  
Indeed, for a homogeneous dusty HII region it is well known that only 
a negligible fraction of the Ly$\alpha$ photons can escape the medium 
(c.f.\ Spitzer 1978).  Observations by e.g.\ Hartmann, Huchra, \& 
Geller (1984) and Terlevich et al.\ (1993) have firmly established the 
strong decrease in Ly$\alpha$ equivalent width with increasing oxygen 
abundance.

The presence of dust in galactic HII regions therefore needs to be 
studied in detail.  In a cosmological context, the formation of dust 
requires the presence of metals, and is therefore intimately related 
to the star formation process itself (Miralda-Escud\'e \& Rees 
1998).  Connected to this metal production is the question of the 
early enrichment of the intergalactic medium (IGM) as seen in damped 
Ly$\alpha$ systems, as well as that of the hot cluster gas observed in 
lower redshift systems.  The interplay between the IGM and the 
galactic environment likely plays an important role in the 
distribution of metals and dust through galactic wind expulsion and 
accretion (Ferrara 1998).  These effects are particularly important in 
the light of recent observations by Pettini et al.\ (1998).
Although their signal--to--noise ratio should be improved, these
observations suggest that damped Ly$\alpha$ systems do not trace 
compact dense regions, i.e.\ primordial galaxies, but reflect the more 
diffuse IGM enriched by these same primeval structures.  As such, the 
metal content of a galaxy or the IGM may show two quite different sides 
of the same star formation process.

The importance of dust has been encountered in the past both 
observationally and theoretically.  Obscuration effects have been 
studied for quasars by Fall \& Pei (1993) and galaxies by Charlot \& 
Fall (1993).  The latter authors perform a study somewhat similar to 
the present paper, but here we have included the effects of inhomogeneity 
and large amounts of dust, and also utilized the physically motivated
Press--Schechter formalism for the star--formation history.    Local 
starbursts are known to emit the bulk of their energy in the infrared 
through dust absorption and re--emission.  The evolution of the universal 
average star--formation rate (Madau et al.\ 1996) must likely be modified 
for dust extinction to reflect the true star formation history of the 
universe.  Finally, recent SCUBA (JCMT) results indicate the presence 
of extremely dusty galaxies (Smail et al.\ 1998) at high redshifts 
($z\sim3$), which could also be related to the production of the 
recently discovered cosmic infrared background radiation (Puget et al.\ 1996; 
Hauser et al.\ 1998).

The aim of this paper is to investigate the importance of dust 
for the recently discovered Ly$\alpha$ emitters; to show that it is 
possible to derive constraints from the observational data on the 
formation of galactic structure, and on the first generation of stars; 
and finally to make predictions for higher, previously unexplored 
redshifts.  It is assumed that the bulk of the Ly$\alpha$ photon 
production is driven by stars, and that the contribution of type II 
supernova shocks or non--thermal emission is negligible (Charlot \& Fall 
1993).  An issue which will not be addressed here in detail is the 
formation of Ly$\alpha$ galaxies (e.g.\ Primack et al.\ 1998).  
Although this is clearly an important question, the aim here is to set 
up a general cosmological model, based on the Press--Schechter 
formalism, where the absolute star formation rate and the physical 
state of the system are varied while the observational consequences 
are explored.

\section{Ly$\alpha$ Emission from HII Regions}

It is well known that even a modest amount of dust inside an HII 
region is sufficient to attenuate all of the produced Ly$\alpha$ 
radiation, because the line optical depth is of the order of $10^4$.  
However, this situation can be alleviated if the dust content is 
negligible, or, more interestingly, if the medium is inhomogeneous 
(Neufeld 1991).  The latter author has argued that in a multi--phase 
medium the escape of Ly$\alpha$ radiation is significantly enhanced 
compared to the standard case of resonant scattering in a homogeneous 
medium.

The radiative transfer problem which needs to be solved for the 
Ly$\alpha$ line is well studied and has been investigated extensively 
(e.g.\ Adams 1972; Hummer \& Kunasz 1980; Neufeld 1990).  Our aim here 
is to model an individual galaxy with a range of masses for the 
ionizing stars, dust content, and inhomogeneity.  A numerical Monte 
Carlo approach is therefore adopted whose general implementation can 
be found in Spaans (1996; and references therein).  In these 
computations, a Scalo IMF is assumed for the spectral types of the 
central stars in the HII regions, ranging from O5 to B1.  The stars 
are distributed in a statistically homogeneous manner inside a 
percolating multi--phase medium.  A typical line width $\Delta V$ of 8
km/s has been used.  Care was taken to sample the line profile 
sufficiently far into the wings, more than $10^2(N_{20}\Delta 
V_5)^{1/3}$ km s$^{-1}$ for a homogeneous medium with a hydrogen 
column density $N_{20}$ in units of $10^{20}$ cm$^{-2}$ and a velocity 
dispersion $\Delta V_5$ in units of $10^5$ cm s$^{-1}$ (Harrington 
1973).  For an inhomogeneous medium the formalism of Neufeld (1991) 
for a multi--phase medium is adopted.

\begin{figure}[t]
\includegraphics{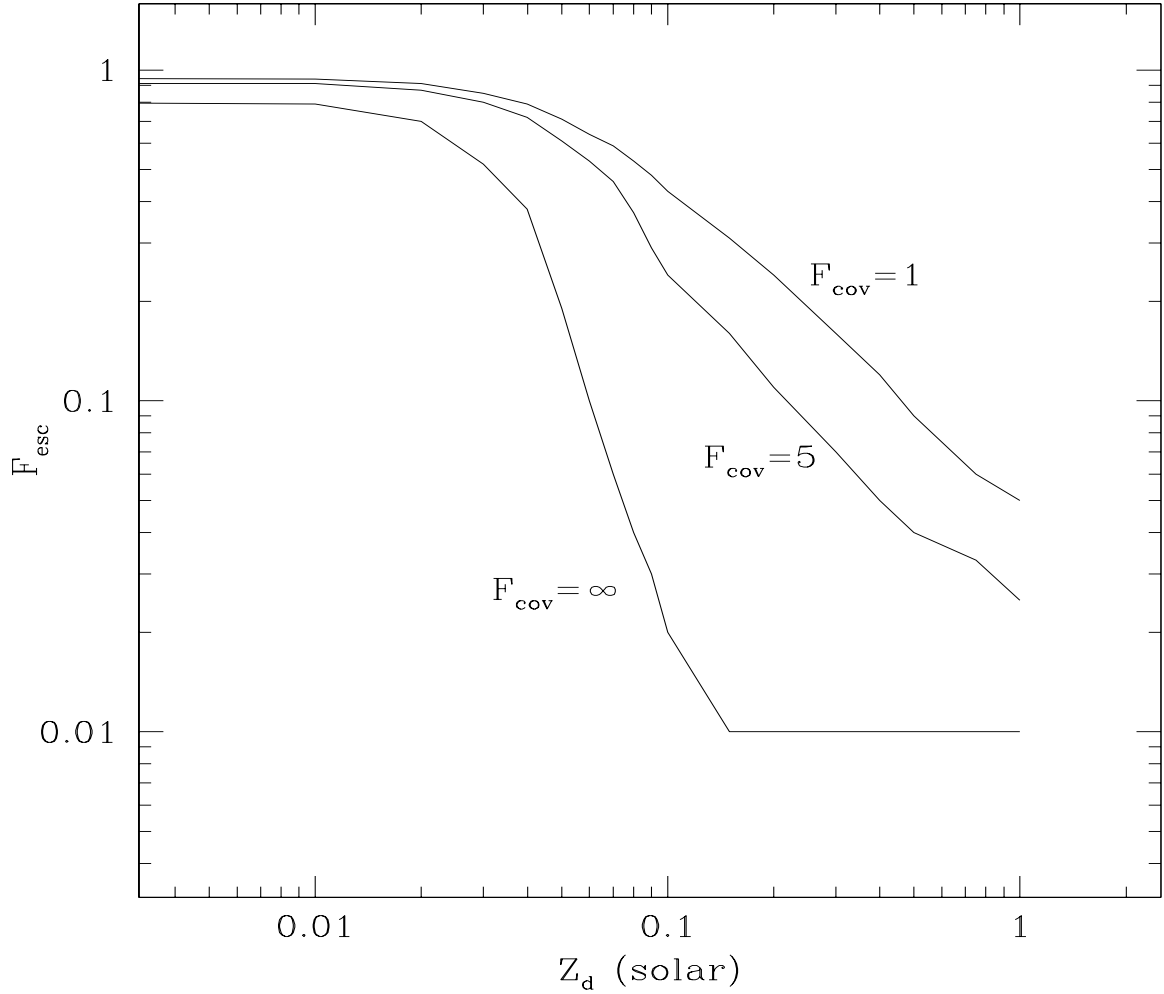}
\vspace*{4.5in}
\caption[Ly$\alpha$ Escape Fractions]
{\label{fig:fesc} The escape fraction $F_{\rm esc}$ of Ly$\alpha$ 
photons from an inhomogeneous medium, as a function of dust content.  
The three curves correspond to three different values of the covering 
factor of opaque clumps, $F_{\rm cov}=$1, 5, and $\infty$.}
\end{figure}

Analogously to the Neufeld investigation, we parameterize the
multi--phase medium by opaque clumps embedded in an inter--clump medium
of negligible opacity.  The clump covering factor $F_{\rm cov}$, i.e.\
the average number of clumps along a line of sight, then fixes the
degree of inhomogeneity.  For the work presented here, the escape
fractions $F_{\rm esc}=\exp(-\tau_*)$ were computed on a grid of
models with various dust contents, $Z_{\rm d}=10^{-2}-1$ solar (Draine
\& Lee 1984), and covering factors, $F_{\rm cov}=1,5,\infty$.  The
results of the numerical computations for these three covering factors
are shown in Figure~\ref{fig:fesc}.  As this figure shows, the
covering factor has a significant effect on the escape fraction.  In
particular, the inhomogeneous percolating slabs are much more
transparent than the homogeneous ones; the difference around $Z_{\rm
d}\sim10^{-1}$ solar is over an order of magnitude.  The reason for
the large difference is that photons incident upon highly opaque
clumps will be reflected back into the interclump medium.  The photons
do not penetrate the opaque clumps and spend most of their time in the 
interclump medium, where the opacity is very small, facilitating their 
escape through repeated reflections off the clumps (Neufeld 1991).

In order to check the accuracy
of our computations, we performed similar calculations for the slab 
geometry adopted by Neufeld (1991).  We were able to obtain good 
agreement with the analytical solutions for these geometries.  In particular, 
the result for the effective optical depth $\tau_*\approx (\Delta V/\Delta 
V_*)^{1/2}$ with 
\begin{equation}
\Delta V_*=680 F_{\rm cov}^{-2}Z_{\rm d}^{-1/2}x({\rm H}^0),
\end{equation}
was reproduced with high accuracy for 
the neutral hydrogen abundance $x$.  For computational convenience, 
the numerical calculations were terminated once the escape fraction 
reached $10^{-2}$, although the results presented below do not depend 
on this choice.  In all of the calculations below, we assume the medium
to be homogeneous until its metallicity exceeds 3\% solar, and allow
for inhomogeneities once this level of enrichment is reached.

\section{Cosmological Abundance of Ly$\alpha$ Emitters}

In order to model the cosmological abundance of high--redshift 
Ly$\alpha$ emitters as a function of redshift, we assume that the 
formation of dark matter halos follows the Press--Schechter (1974) 
theory.  Hence, the net rate of change in the comoving number density 
of halos with mass $M_{\rm halo}$ is given by the derivative ${d\over 
dz}(dn_{\rm c}/dM)$, where $dn_{\rm c}/dM$ is the Press--Schechter 
mass function (comoving number density per unit mass).  This rate 
includes a negative contribution from the disappearance of small halos 
in merger events; the net ${d\over dz}(dn_{\rm c}/dM)$ becomes 
negative when mergers dominate, i.e.\ for masses below a characteristic 
mass scale $M_*(z)$ at any given redshift $z$.  We make the 
simplifying assumption that no halos form with masses below $M_{\rm 
halo}<M_*(z)$, and therefore set ${d\over dz}(dn_{\rm c}/dM)=0$ for 
halos below this mass.

Next, we assume that every halo forms a galaxy that goes through a 
Ly$\alpha$ emitting phase.  Indeed, after a halo collapses and 
virializes, its gas can fragment into stars, provided it can cool 
efficiently below its initial virial temperature.  Because of the lack 
of a significant amount of $\rm H_{2}$ molecules at high redshifts 
(Haiman, Rees \& Loeb 1997), the leading cooling mechanism is expected 
to be collisional excitation of atomic hydrogen.  The corresponding 
requirement for halos to cool efficiently is that their mass is 
at least $M_{\rm min}\sim10^{8}{\rm M_\odot}[(1+z)/11]^{-3/2}$ (Haiman 
\& Loeb 1998).  We therefore assume that in halos that exceed this 
mass, a fraction $\epsilon_\star$ of the gas turns into stars, with a 
constant rate of star formation over a period of $t_\star$ years.  The 
result of this process is a dwarf galaxy with a stellar mass of 
$M_{\rm star}=\epsilon_\star (\Omega_{\rm b}/\Omega_0) M_{\rm halo}$.  The 
values of $\epsilon_\star$ and $t_{\star}$ could depend on several 
further parameters, such as the formation redshift, halo mass, or the 
initial angular momentum of the gas.  There could be further
complications, such as a time--dependent star formation rate.  A 
treatment of these important issues is beyond the scope of this paper, 
and for simplicity we adopt the simplest assumption, 
i.e.\ that $\epsilon_\star$ and $t_{\star}$ both have the same 
constant values in each halo.
 
Finally, an important effect we must include is an external feedback 
from photo--ionization by the UV background, which heats the gas before 
it is able to cool and condense inside the dark matter potential 
wells.  This feedback is an inevitable consequence of the background 
UV flux that builds up after the re--ionization epoch, when the 
cosmological HII regions have overlapped and the universe is 
transparent to the ionizing flux from each individual source.  Several 
authors have discussed the consequences of this feedback from first 
principles (Babul \& Rees 1992; Efstathiou 1992; Quinn, Katz \& Efstathiou 
1996; Thoul \& Weinberg 1996; Navarro \& Steinmetz 1997), and concluded 
that the collapse of gas is inhibited in halos with circular velocities 
below $v_{\rm circ}=30-80~{\rm km~s^{-1}}$.  The apparent lack of 
high--redshift quasar candidates in the Hubble Deep Field (HDF) also 
requires the existence of some type of feedback of a similar magnitude.
To be consistent with the HDF data, Haiman, Madau \& Loeb (1998) have
found that either the formation, or the fueling of black holes was
prevented in halos with $v_{\rm circ}<75~{\rm km~s^{-1}}$.  Accordingly, 
we impose a minimum circular velocity $v_{\rm circ}=75~{\rm 
km~s^{-1}}$ for the halos of luminous galaxies that form after the 
re--ionization epoch, here taken to be $z_{\rm reion}=10$ 
(cf.\ Haiman \& Loeb 1997).  This constraint results in a minimum 
halo mass of $M_{\rm min}\sim10^{10}{\rm M_\odot}[(1+z)/11]^{-3/2}$,
two orders of magnitude larger than the minimum mass obtained from the 
cooling argument above.

We note that recently discovered high--redshift galaxies (Steidel et
al.\ 1998) have revealed Ly$\alpha$ emission only in a fraction of all
the sources.  This may be consistent with the small duty--cycle of the
Ly$\alpha$ emitting phase considered here; alternatively, $t_\star$
could be increased somewhat if only a fraction $f$ of all halos would
undergo a Ly$\alpha$ emitting phase.  In order to keep the predicted
number density constant, $t_\star$ would then need to be increased
approximately to $\sim t_\star/f$.  The Press--Schechter halo
formation rate also depends on the cosmology and power spectrum.  For
our ``standard model'' we have adopted the concordance model of
Ostriker \& Steinhardt (1995), i.e.\ a flat $\Lambda$CDM model with a
slightly tilted power spectrum ($\Omega_0,\Omega_\Lambda, \Omega_{\rm
b},h,\sigma_{8h^{-1}},n$)=(0.35, 0.65, 0.04, 0.65, 0.87, 0.96).
Convenient expressions for the differential volume element, luminosity
distance, and time--redshift relation in this model are given in terms
of elliptic integrals by Eisenstein (1997); a useful fitting formula
for the growth function is given by Carroll~et~al.~(1992).  In
addition to this $\Lambda$ cosmology, below we consider an open model
with the same parameters, except the power spectrum normalization is
then changed to $\sigma_{8h^{-1}}=1.15$.

\section{Model Description}

In order to compare our models directly with observations, we need to 
compute the surface density of Ly$\alpha$ emitters on the sky above a 
given observational threshold flux.  As discussed above, each of our 
model galaxies undergoes continuous star formation for $t_\star$ 
years.  During this time--interval, we relate the local star formation 
rate (SFR) to an intrinsic Ly$\alpha$ luminosity, assuming case B 
recombination, and using Kennicutt's (1983) relation between the SFR 
and Ly$\alpha$ luminosity, resulting in $L_{\alpha,{\rm 
int}}=(\dot{M}/{\rm M_\odot~yr^{-1}})~10^{42}~{\rm erg~s^{-1}}$.  Here 
$\dot{M}\equiv\Omega_{\rm b}/\Omega_0 M_{\rm halo} \epsilon_\star / 
t_\star$ is the adopted star formation rate.  We assume that each model
galaxy produces stellar dust during the same time--interval, and that 
some fraction $x_g$ of this dust stays within the galaxy and enriches 
its interstellar medium, while the remaining fraction $1-x_g$ is blown 
out of the galaxy by supernovae, stellar winds, or as a result of 
mergers between galaxies (Gnedin 1998).  Galaxies that form 
subsequently will be assembled from an intergalactic gas that is 
pre--enriched with this ejected dust.

\begin{figure}[t]
\includegraphics{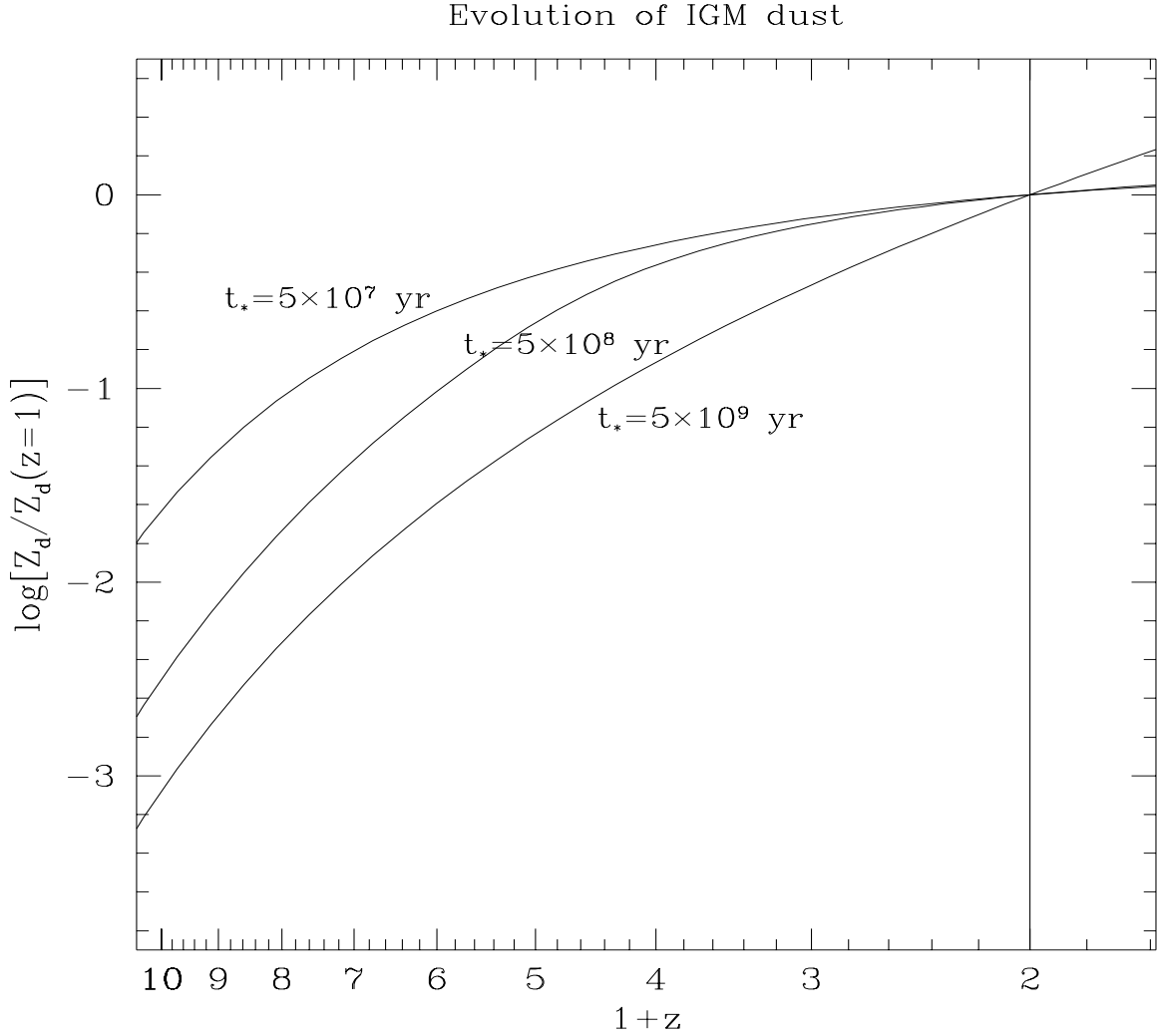}
\vspace*{4.5in}
\caption[Evolution of IGM Metallicity]
{\label{fig:zigm} The derived evolution of the IGM dust content, 
$Z_{\rm d, IGM}$ in our models.  The three curves correspond to three 
different values of $t_\star$.  Since $\epsilon_\star$ is kept fixed, 
the assumed star formation rate in each individual galaxy scales as 
$1/t_\star$.}
\end{figure}

It is difficult to estimate from first principles either the fraction 
$x_g$, or the size of the region around each galaxy polluted by the 
blown--out dust (although see recent 3--D numerical simulations by 
MacLow \& Ferrara 1998).  Instead, here we adopt a phenomenological 
approach, and introduce two additional parameters that characterize 
the overall dust content of galaxies.  The first parameter, $Z_{\rm d, 
ISM}(t)$ is the mass fraction of dust (in solar units) in the
interstellar medium of each galaxy, due only to the self--enrichment 
from the retained dust fraction $x_g$.  Consistent with our assumption 
of a constant star formation rate, we assume that the amount of 
retained dust increases linearly with time for $t_\star$ years, until 
it reaches the final value of $Z_{\rm d,ISM}(t_\star)$.  Similarly, we 
assume that each galaxy deposits dust into the surrounding IGM at 
a constant rate for $t_\star$ years, and introduce a second parameter, 
$Z_{\rm d, IGM}(z)$, denoting the resulting redshift--dependent 
average dust content of the {\it polluted regions} within the intergalactic 
medium.  Note that the ejected dust may not be fully mixed into the 
IGM, and the universal average dust fraction of the IGM could be lower 
than $Z_{\rm d, IGM}$.  Here we avoid this issue by defining $Z_{\rm 
d,IGM}(z)$ to denote the average dust content only within the pockets 
of intergalactic gas from which subsequent galaxies form.  
Observationally, the metallicity of gas in galaxy clusters upto 
redshift $z=1$ is approximately 1/3 solar, which we take as the 
representative value for $Z_{\rm d, IGM}(z=1)$.  Figure~\ref{fig:zigm} 
shows the evolution of $Z_{\rm d, IGM}(z)$ in our models, obtained by 
summing over the dust output of all Press--Schechter halos, and 
assuming three different values of $t_\star$.  Note that the longer
the dust--producing phase in each individual galaxy, the steeper 
the buildup to the average $Z_{\rm d, IGM}(z=1)$ has to be, resulting
in a lower dust content at higher redshifts.

In summary, a galaxy that forms at redshift $z_f$ has, at a later 
redshift $z$, a total dust content of $Z_{\rm d, IGM}(z)+Z_{\rm d, 
ISM}(\Delta t)$, where $\Delta t$ is the time elapsed from $z_f$ to 
$z$.  Both types of dust contribute to reduce the intrinsic Ly$\alpha$ 
luminosity $L_{\alpha,{\rm int}}$ implied by the star--formation rate.  This 
reduction is expressed by the escape fraction as discussed in Section 
2 above.  Apart from the underlying cosmology, our model therefore 
has five adjustable parameters, $t_\star$, $F_{\rm cov}$, 
$\epsilon_\star$, $Z_{\rm d, IGM}(z=1)$, and $Z_{\rm d, 
ISM}(t_\star)$, that uniquely determine the number density of 
Ly$\alpha$ emitters at any flux and redshift.  More precisely, at 
redshift $z$, the comoving number density $n_c(z,F_\alpha)$ of 
emitters with observed line flux between $F_\alpha$ and 
$F_\alpha+dF_\alpha$ is given by a sum over halos of different 
formation redshifts and ages that exist at redshift $z$,
\begin{equation}
n_{\rm c}(z,F_\alpha)=\int_z^{\infty} dz^\prime
\frac{dM_{\rm halo}}{dF_\alpha}(z,z^{\prime},F_\alpha) 
\left.\frac{d^2n_{\rm c}}{dM dz^\prime} 
\right|_{M_{\rm halo}(z,z^{\prime},F_\alpha),z^\prime},
\label{eq:ncom} 
\end{equation}
where the factor $dM_{\rm halo}/dF_\alpha$ converts the number density 
per unit mass interval to number density per unit line flux interval.  
The line intensity $F_\alpha$ is given by $F_\alpha=F_{\rm 
esc}(z,z^\prime) L_{\alpha,{\rm int}}(M_{\rm halo},\epsilon_\star, 
t_\star)/4\pi d_L(z)^2$, where $d_L(z)$ is the standard cosmological 
luminosity distance.

\section{Results and Discussion}

We first define a ``standard model'' by the set of parameter values 
$t_\star=5\times10^8~{\rm yrs}$, $F_{\rm cov}=5$, $\epsilon_\star=10\%$, 
and $Z_{\rm d,IGM}(z=1)=Z_{\rm d, ISM}(t_\star)=0.3~{\rm Z_\odot}$. 
The first three of these values are broadly consistent with the star 
formation rates, covering factors, and stellar mass fractions 
estimated in low--redshift galaxies; the latter two are chosen based 
on the metallicity of gas in galaxy clusters and within the 
interstellar medium of low--redshift galaxies (c.f.\ Mushotzky \&
Loewenstein 1997; Lada, Evans, \& Falgarone 1997; Ho, Filippenko, \&
Sargent 1997; Hammer et al.\ 1997; Young et al.\ 1996).
In Table~\ref{tab:models} we summarize the parameters of our standard 
model, as well the ranges we have considered for each parameter.

\begin{table}[b]
\caption{\label{tab:models} 
The assumed parameter values in our standard model and its variants.}
\vspace{0.3cm}
\begin{center}
\begin{tabular}{|c||c|c|}
\hline
Parameter & Standard Model  & Range Considered \\
\hline
\hline  
$t_\star$            & $5\times10^8~{\rm yrs}$     & $5\times10^{7-9}~{\rm yrs}$ \\
\hline
$F_{\rm cov}$        & $5$                         & $1-\infty$  \\
\hline
$\epsilon_\star$     & $10\%  $                    & $2-20\%$ \\
\hline  
$Z_{\rm d, IGM}(z=1)$   & $0.3~{\rm Z_\odot}$      & $0.1-1~{\rm Z_\odot}$ \\
\hline
$Z_{\rm d, ISM}(t_\star)$ & $0.3~{\rm Z_\odot}$    & $0.1-1~{\rm Z_\odot}$ \\
\hline
\hline  
\end{tabular}
\end{center}
\end{table}

\begin{figure}[t]
\includegraphics{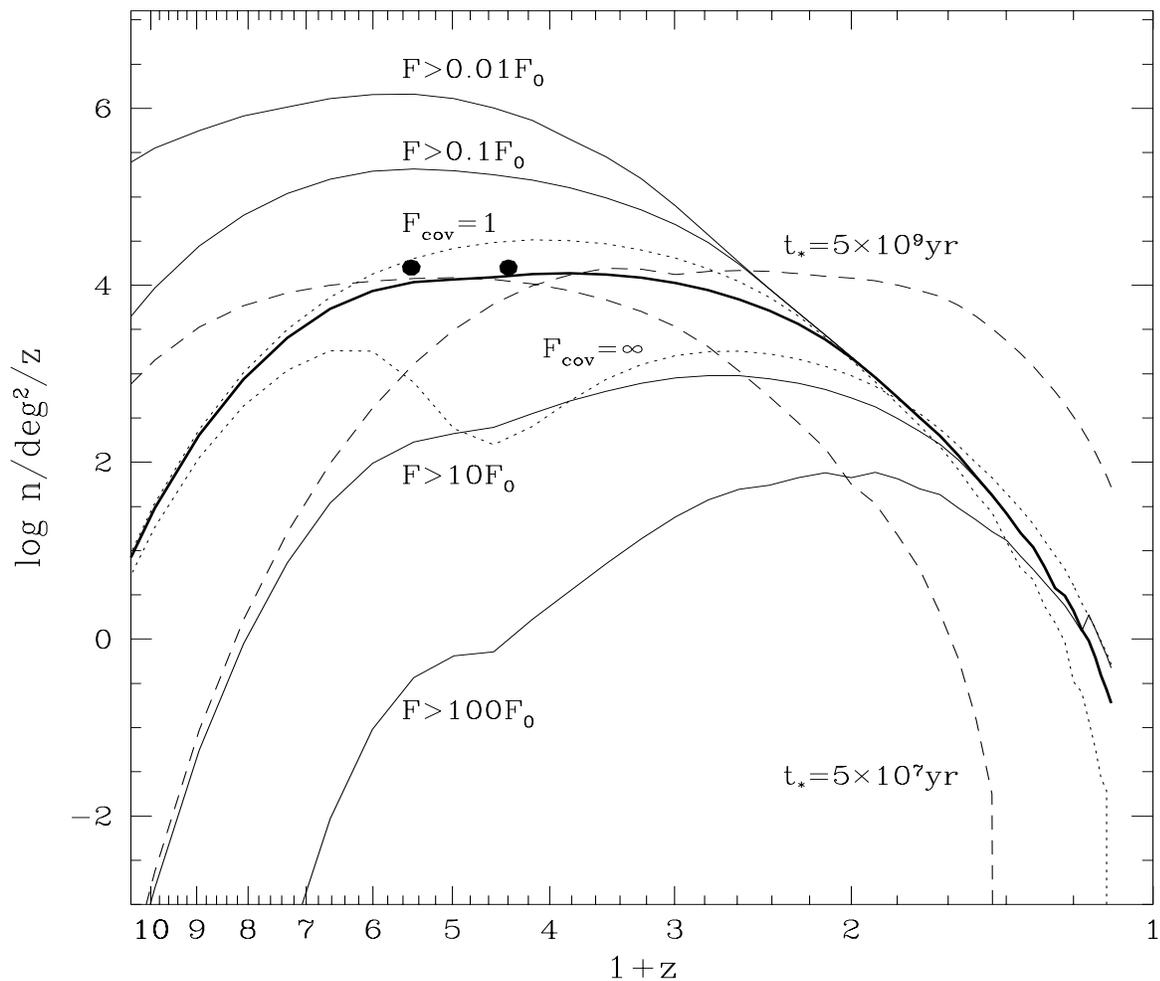}
\vspace*{4.5in}
\caption[Density of Ly$\alpha$ Emitters]
{\label{fig:res1} The surface density of Ly$\alpha$ emitters in our 
standard model (solid lines) with fluxes above different values of the 
detection threshold.  The two data points are taken from Hu et al.\
(1998).  For the fixed threshold $F_0=1.5\times10^{-17}~{\rm 
erg~cm^{-1}~s^{-1}}$, the dashed lines show how the surface density 
changes if the assumed star--formation rate is increased or decreased 
by a factor of 10.  Similarly, the dotted lines show the surface 
density when the covering factor is changed to $F_{\rm cov}=1$, or 
$\infty$.}
\end{figure}

In Figure~\ref{fig:res1} we show the resulting surface density of
Ly$\alpha$ emitters in our standard model, with fluxes above different
values of the detection threshold.  We have chosen the quoted
5$\sigma$ detection limit of the recent narrow--band survey of Hu et
al.\ (1998), $F_0=1.5\times10^{-17}~{\rm erg~cm^{-1}~s^{-1}}$, as a
fiducial threshold value.  We also indicate the results from this
survey for the surface density of emitters at the two redshifts
$z=3.4$ and $z=4.5$.  As the figure shows, the model is in good
agreement with these two available data points.  Although the
agreement is encouraging, it is important to examine the sensitivity
of this result to changes in the model parameters.  In
Figure~\ref{fig:res1}, we demonstrate the effect of changing the star
formation rate by changing $t_\star$, while leaving $\epsilon_\star$
fixed.  This corresponds to changing the SFR but not the total stellar
mass.  The dashed lines show that when $t_\star$ is decreased by a
factor of 10, then both starlight and dust is produced earlier, i.e.\
sooner after the formation of each galaxy.  This causes the abundance
of emitters to peak at higher redshifts, but this makes the agreement
with the data points only slightly worse.  On the other hand, when
$t_\star$ is increased by a factor of 10, the abundance of emitters
peaks at lower redshifts, and the discrepancy with the data points
becomes significant: the model now underpredicts the abundance by more
then an order of magnitude at $z=4.5$.  Finally, we show in
Figure~\ref{fig:res1} the effect of changing the clump covering factor
$F_{\rm cov}$.  The dotted lines show that when the medium is assumed
to be inhomogeneous with $F_{\rm cov}=1$, the model still fits the abundance
at $z=4.5$, but overpredicts it by a factor of $\sim2$ at $z=3.4$.  On
the other hand, for a homogeneous medium ($F_{\rm cov}=\infty$), the
escape fraction drops dramatically (cf.\ Fig.~\ref{fig:fesc}),
resulting in a substantial decrease in the surface density of
emitters, again leading to a significant (two orders of magnitude)
discrepancy with the Hu et al.\ data at $z=3.4$.

It is interesting to note that the surface density of emitters is a steep
function of the detection threshold.  In particular, Figure~\ref{fig:res1}
reveals that, depending on the values of the other parameters, raising the flux
threshold by an order of magnitude can decrease the predicted number of
emitters by 2--3 orders of magnitude in the relevant redshift range, $z=2-5$.
This is an important feature of the present models, when we consider the
existing data together with the lack of detections of Ly$\alpha$ emitters in
previous surveys.  In particular, based on a lack of detections, Thompson,
Djorgovski \& Trauger (1995) have deduced a limit on
the comoving volume density of emitters that is a steep function
of the detection threshold. The surface density found by Hu et
al.\ (1998) is barely consistent with this limit at the faint end of the
threshold (${\rm L_{Ly\alpha}\sim10^{42}~erg~s^{-1}}$).  Consistency with the
Thompson, Djorgovski \& Trauger (1995) limits then requires that the model
predicts a decline in the surface density with increasing flux threshold, and
that this decline is at least as steep as the inverse square of the threshold.
As Figure~\ref{fig:res1} shows, our models indeed have this sharply
declining feature in the redshift range $2.78<z<4.89$ of Thompson, Djorgovski
\& Trauger (1995), and can therefore simultaneously explain the new
detections without violating the existing upper limits on the abundance
of emitters.

\begin{figure}[t]
\includegraphics{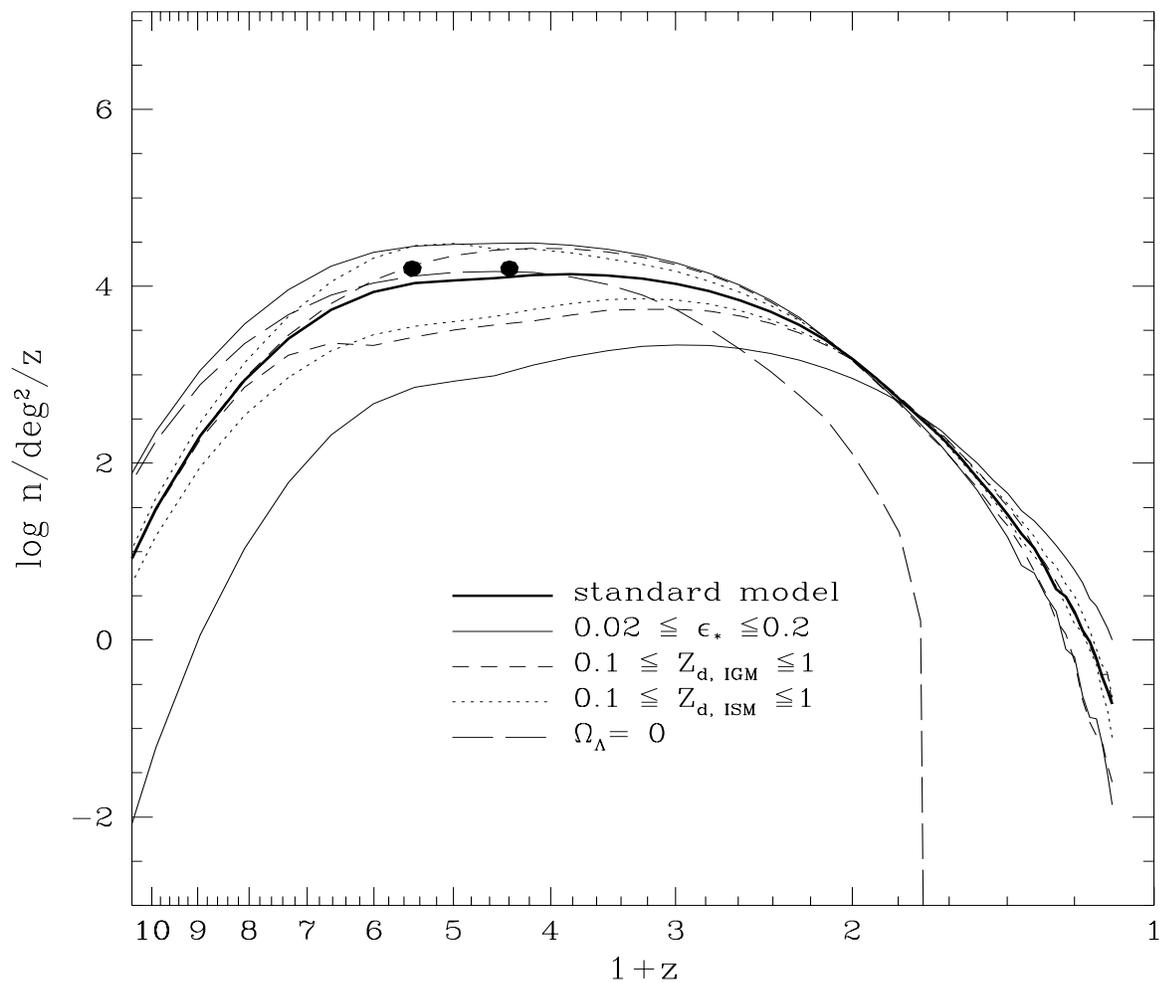}
\vspace*{4.5in}
\caption[Density of Ly$\alpha$ Emitters] {\label{fig:res2} Same as
Fig.~\ref{fig:res1}, but now the solid lines show how the surface
density changes with the star formation efficiency, $\epsilon_\star$;
the dashed and dotted lines show the effect of changing the assumed
IGM or ISM dust contents in the interval $0.1-1~{\rm Z_\odot}$; and
the long--dashed lines show the effect of changing the cosmology.

}
\end{figure}

Finally, in Figure~\ref{fig:res2}, we demonstrate the effects of
changing the cosmology, the star formation efficiency
$\epsilon_\star$, or the intergalactic and galactic dust contents
$Z_{\rm d, IGM}$ and $Z_{\rm d, ISM}$.  In general, lowering the
amount of dust raises the escape fraction, and the Ly$\alpha$ line
luminosity, and therefore increases the number of emitters above a
fixed threshold, and vice versa.  However, we find our results to be
less sensitive to these changes than those in the star formation rate
or in the clump covering factor, shown in Figure~\ref{fig:res1}.  In
particular, when either $Z_{\rm d, IGM}$ (dashed lines) or $Z_{\rm d,
ISM}$ (dotted lines) is varied between 0.1 and 1 ${\rm Z_\odot}$, the
corresponding range in the surface density of emitters is only about
an order of magnitude at $z\sim4$, and even less at higher and lower
redshifts.  On the other hand, the results are more sensitive to
changes in the star formation efficiency (solid lines). For the range
$0.02\leq\epsilon_\star\leq0.2$, the abundance varies by almost two
orders of magnitude.  Note that a low star formation efficiency
results in an underestimate of the abundance that can not be
compensated by changes in the other parameters
(cf. Fig.~\ref{fig:res1}).  Finally, the long--dashed line in
Fig.~\ref{fig:res2} shows the effect of changing the cosmology 
to an open CDM model.  As the figure shows, around redshifts
$z=3-5$, the two cosmological models give almost identical results
(although they differ at lower redshifts). We conclude that the
parameters that are best constrained by the data are $t_\star$,
$F_{\rm cov}$, and $\epsilon_\star$.

\section{Conclusions}

We have studied the abundance of high--redshift Ly$\alpha$ emitters by
combining the Press--Schechter theory, describing the cosmological
formation of halos, with a simple prescription for an inhomogeneous
dust distribution and its dependence on the star formation process.
We find that these simple models can explain reasonably well the
recent observations and earlier non--detections of high--redshift
Ly$\alpha$ emitters.  We find that it is possible to place significant
constraints on the inhomogeneity of the spatial distribution of dust,
parameterized by the clump covering factor $F_{\rm cov}$, the star
formation rate, parameterized by the duration of the Ly$\alpha$
emitting phase $t_\star$, and the overall star formation efficiency,
$\epsilon_\star$.  Our results show that in order to reproduce the
surface density observed by Hu et al.\ (1998), $F_{\rm cov}$ must be
of order unity, $t_\star$ must not exceed $\sim5\times10^8$ yr, and
the star formation efficiency must be at least $\sim5\%$.  These
numbers should be predicted by more complete and detailed models of
galactic evolution, and will be useful discriminators between such
models.

Our models also predict that more Ly$\alpha$ galaxies will be
detectable, around the present flux threshold, upto redshifts as high 
as $\sim$8, indicating that for a simple Press--Schechter cosmological 
model there can be significant Ly$\alpha$ activity in the early phases 
of galaxy evolution.  This is especially interesting in the light of 
recent discoveries of Ly$\alpha$ galaxy candidates at redshifts $z>5$.  
Three spectroscopically confirmed objects have been reported by Hu et 
al.\ (1998) at a redshift of 5.64, Weymann et al.\ (1998) at a 
redshift of 5.60, and Dey et al.\ (1998) at a redshift of 5.34.  
Clearly, the search for Ly$\alpha$ emitters is beginning to enjoy 
considerable successes and these primeval objects may turn out to hold 
important clues both for galaxy evolution, as well as ISM physics.

\acknowledgements

We thank Arif Babul for useful comments. ZH was supported by the 
DOE and the NASA grant NAG 5-7092 at Fermilab.  M.S. has been 
supported by NASA through Hubble Fellowship grant \#~HF-01101.01-97A, 
awarded by the Space Telescope Science Institute, which is operated 
by the Association of Universities for Research in Astronomy, Inc., 
for NASA under contract NAS 5-26555.
 
\clearpage
 
{\StartRef

\Ref  Adams, T.F.\ 1972, ApJ, 174, 439

\Ref Babul, A., \& Rees, M. J. 1992, MNRAS, 255, 346

\Ref Carroll, S. M., Press, W. H., Turner, E. L. 1992, ARA\&A, 30, 499

\Ref Chambers, K., \& Miley, G.K.\ 1990, in The Evolution of the Universe of Galaxies: The Edwin Hubble Centennial Symposium, ed.\ R.\ Kron (ASP Conf. Ser., 10), p.\ 373

\Ref Charlot, S., \& Fall, S.M.\ 1993, ApJ, 415, 580

\Ref Dey, A., Spinrad, H., Stern, D., Graham, J.R., \& Chaffee, F.H.\ 1998, ApJ, 498, L93

\Ref Djorgovski, S.\ 1992, in Cosmology and Large Scale Structure in the Universe, ASP Conf.\ Ser., 24, ed.\ R.R.\ de Carvalho (San Francisco, ASP), p.\ 73

\Ref Draine, B.T., \& Lee, H.M.\ 1984, ApJ, 285, 89

\Ref Efstathiou, G.\ 1992, MNRAS, 256, 43

\Ref Eisenstein, D.J.\ 1997, ApJ, submitted, preprint astro-ph/9709054

\Ref Fall, S.M., \& Pei, Y.C.\ 1993, ApJ, 402, 479

\Ref Ferrara, A.\ 1998, ApJ, 499, L17

\Ref Gnedin, N.Y.\ 1998, MNRAS, 294, 407

\Ref Haiman, Z., \& Loeb, A.\ 1997, ApJ, 483, 21

\Ref Haiman, Z., \& Loeb, A.\ 1998, ApJ, 503, 505

\Ref Haiman, Z., Madau, P., \& Loeb, A.\ 1998, ApJ, in press, astro-ph/9805258

\Ref Haiman, Z., Rees, M.J., \& Loeb, A.\ 1997, ApJ, 476, 458 

\Ref Hammer, F., Flores, H., Lilly, S.J., Crampton, D., Le Fevre, O.,
Rola, C., Mallen-Ornelas, G., Schade, D., \& Tresse, L.\ 1997, ApJ, 481, 49

\Ref Harrington, J.P.\ 1973, MNRAS, 162, 43

\Ref Hartmann, L.W., Huchra, J.P., \& Geller, M.J.\ 1983, ApJ, 287, 487

\Ref Hauser, M., et al.\ 1998, ApJ, in press, astro-ph/98061667

\Ref Ho, L.C., Filippenko, A.V., \& Sargent, W.L.W.\ 1997, ApJ, 487, 579

\Ref Hu, E.M., Cowie, L.L., \& McMahon, R.G.\ 1998, astro-ph/9803011

\Ref Hummer, D.G., \& Kunasz, P.B.\ 1980, ApJ, 236, 609

\Ref Kennicutt, R. C., Jr.\ 1983, ApJ, 272, 54

\Ref Lada, E.A., Evan, N.J.\ II, \& Falgarone, E.\ 1997, ApJ, 488, 286

\Ref MacLow, M., \& Ferrara, A.\ 1998, astro-ph/9801237

\Ref Madau, P., Ferguson, H.C., Dickinson, M.E., Giavalisco, M., Steidel, C.C., \& Fruchter, A.\ 1996, MNRAS, 283, 1388

\Ref Meier, D.\ 1976, ApJ, 207, 343

\Ref Miralda-Escud\'e, J., \& Rees, M.J.\ 1998, ApJ, 497, 21

\Ref Mushotzky, R.F., \& Loewenstein, M.\ 1997, ApJ, 481, L63

\Ref Navarro, J. F., \& Steinmetz, M.\ 1997, ApJ, 478, 13

\Ref Neufeld, D.A.\ 1991, ApJ, 370, L85

\Ref Neufeld, D.A.\ 1990, ApJ, 350, 216

\Ref Ostriker, J. P., \& Steinhardt, P.J.\ 1995, Nature, 377, 600

\Ref Partridge, R.B., \& Peebles, P.J.E.\ 1967, ApJ, 147, 868

\Ref Pettini, M., Ellison, S.L., Steidel, C.C., \& Bowen, D.V.\ 1998, astro-ph/9808017

\Ref Press, W.H., \& Schechter, P. L.\ 1974, ApJ, 181, 425

\Ref Primack, J.R., Somerville, R.S., Faber, S.M., \& Wechsler, R.H.\ 1998, astro-ph/9806263

\Ref Pritchet, C.J.\ 1994, Publ.\ Astron.\ Soc.\ Pac., 106, 1052

\Ref Puget, J.-P., Abergel, A., Bernard, J.-P., Boulanger, F., Burton,
W.B., D\'esert, F.-X., Hartmann, D.\ 1996, A\&A, 308, L5

\Ref Quinn, T., Katz, N., \& Efstathiou, G.\ 1996, MNRAS, 278, 49

\Ref Smail, I., Ivison, R.J., Blain, A.W., \& Kneib, J.-P.\ 1998, astro-ph/9806061

\Ref Spaans, M.\ 1996, A\&A, 307, 271

\Ref Spitzer, L.\ 1978, Physical Processes in the Interstellar Medium, (New York, Wiley)

\Ref Steidel, C.C., et al., 1998, in Proc. of the ESO Symposium ``From Recombination to  Garching'', held Aug. 2-7, 1998, Garching, Germany

\Ref Terlevich, E., Diaz, A.I., Terlevich, R., \& Garcia Vargas, M.L.\ 1993, MNRAS, 260, 3

\Ref Thompson, D., Djorgovski, S., \& Trauger, J.\ 1995, AJ, 110, 963

\Ref Thoul, A.A., \& Weinberg, D.H.\ 1996, ApJ, 465, 608

\Ref Weymann, R.J., Stern, D., Bunker, A., Spinrad, H., Chaffee, F.H., Thompson, R.I., \& Storrie-Lombardi, L.J.\ 1998, astro-ph/9807208

\Ref Young, J.S., Allen, L., Kenney, J.D.P., Lesser, A., \& Rownd, B.\ 1996, AJ, 112, 1903

}

\clearpage
\newpage

\clearpage
\newpage

\clearpage
\newpage

\clearpage
\newpage

\end{document}